\documentclass[twocolumn,showpacs,preprintnumbers,amsmath,amssymb]{revtex4}
\usepackage{graphicx}% Include figure files
\usepackage{dcolumn}% Align table columns on decimal point
\usepackage{bm}% bold math

\begin{document}

%\preprint{APS/123-QED}

\title{Laser frequency locking by direct measurement of detuning}%

\author{A. Ratnapala, C. J. Vale, A. G. White, M. D. Harvey, N. Heckenberg
 and H. Rubinsztein-Dunlop}

\affiliation{School of Physical Sciences, University of
Queensland, St. Lucia 4072, Australia
}%

\date{\today}% It is always \today, today,

\begin{abstract}
We present a new method of laser frequency locking in which the
feedback signal is directly proportional to the detuning from an
atomic transition, even at detunings many times the natural
linewidth of the transition.  Our method is a form of sub-Doppler
polarization spectroscopy, based on measuring two Stokes
parameters ($I_2$ and $I_3$) of light transmitted through a vapor
cell. This extends the linear capture range of the lock loop by up
to an order of magnitude and provides equivalent or improved
frequency discrimination as other commonly used locking
techniques.
\end{abstract}

\maketitle

Laser frequency locking is the process of controlling a laser's
frequency, relative to some physical reference, by means of
feedback.  Various physical references may be used, for example
cavities or atomic transitions \cite{demtroder}, and the methods
described here apply equally well in both cases.  In this work we
focus on the spectroscopy of hyperfine transitions in Rb atomic
vapor which is relevant to work on ultra-cold atoms.

Experiments in atomic physics often require lasers to be locked
precisely to a known atomic transition. Saturated absorption, and
polarization spectroscopy, are two well known techniques which
allow this to be achieved \cite{demtroder}. Both rely on counter
propagating pump and probe beams to measure a single longitudinal
velocity group of atoms within a Doppler broadened profile. In the
case of saturation spectroscopy, the laser may be locked to the
side of an absorption peak \cite{macadam} by direct comparison of
a photodetector signal with a reference voltage, however it is
generally desirable to lock to the top of a peak. This may be
achieved by dithering the laser frequency around a peak and
performing lock-in detection \cite{white}, which provides a
derivative signal of the transmission. This has a zero crossing at
top of the peak which is convenient for locking. Polarization
spectroscopy is somewhat more attractive as it produces a narrow,
sub-Doppler spectrum with a natural zero crossing at resonance
\cite{wieman} without the need for dithering (which can broaden
the laser linewidth) and lock-in detection \cite{hansch}.

Pearman {\it et al.} have studied a form of polarization
spectroscopy locking in detail in \cite{pearman}. Their method
corresponds to a measure of the circular birefringence
(differential refractive index for orthogonal polarizations) of an
atomic sample which is induced by a circularly polarized pump
beam. The locking signal is linear over a range approximately
equal to the power broadened linewidth of the transition
(typically 10\,-\,20\,MHz in rubidium vapors).

The method we present incorporates this along with a measure of
the circular dichroism (differential absorption for two orthogonal
polarizations) which greatly extends the linear region of the lock
signal. Our method stems from the proposal of Harvey and White
\cite{harvey} which is best understood by analyzing the Stokes
parameters of the transmitted probe light.

We wish to quantify the change in the polarization state of an
input laser beam in terms of the anisotropic properties of the
sample.  As the anisotropy is defined in terms of circular
polarizations, we choose to express the total electric field in
the right-left ($R$-$L$) circular basis: $\mathbf{E}=\left[
\begin{smallmatrix} E_R \\ E_L \end{smallmatrix} \right]$ where
$E_R$ and $E_L$ are the right and left circular electric field
components respectively. In this basis the (unnormalized) Stokes
parameters are,
\begin{eqnarray}
    I_0 &=& |E_R|^2 + |E_L|^2, \\
    I_1 &=& 2 |E_R||E_L| \cos{\phi}, \\
    I_2 &=& 2 |E_R||E_L| \sin{\phi}, \\
    I_3 &=& |E_R|^2 - |E_L|^2,
\end{eqnarray}
where $\phi = \phi_R - \phi_L$ is the phase difference between the
right and left circular components. Equations 1\,-\,4 are
equivalent to the familiar expressions in the horizontal-vertical
($H$-$V$) basis, and could also be written in the
diagonal-antidiagonal ($D$-$A$) basis. The normalized Stokes
parameters, $S_i$, are obtained from the ratio $I_i/I_0$. We
recall the physical meaning of these, $I_0$ is the total
irradiance, $I_1$, $I_2$ and $I_3$ are measures of the horizontal,
diagonal and right circular polarizations, respectively
\cite{bornwolf}.

\begin{figure}[h]\centerline{\scalebox{1}
   {\includegraphics{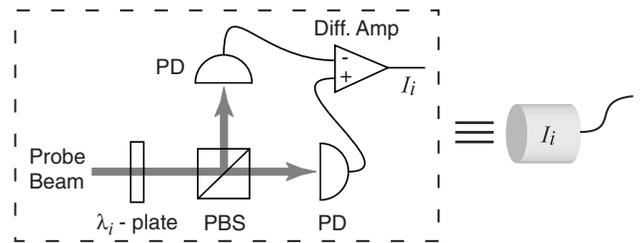}}}
   \caption{Detector for measuring Stokes parameters $I_i$,
   $i$\,=\,1,2\,3. PD = photodetector, PBS = polarizing beam
   splitter, $\lambda_i$-plate = 0, $\lambda/2$, $\lambda/4$
   plate for $i$\,=\,1,2\,3 respectively.}
   \label{fig:detectors}
\end{figure}

With this in mind it is straightforward to design detectors for
each of the Stokes parameters consisting of: for $I_1$ a
polarizing beamsplitter cube and two photodetectors; $I_2$ a
$\lambda/2$ plate and an $I_1$ detector; and $I_3$ a $\lambda/4$
plate and an $I_1$ detector.  Figure \ref{fig:detectors} shows
this schematically. $I_0$ is given by the sum of the two
photodetector signals for any of the above $I_i$ measurements.
In all cases the beamsplitter separates the horizontal and
vertical components, which are then measured and subtracted.  The
waveplates convert ($D$-$A$) to ($H$-$V$) to measure $I_2$ and
($R$-$L$) to ($H$-$V$) to measure $I_3$ respectively.

We return now to the proposal of Harvey and White in which a
measure of the phase difference between two orthogonal field
components, provides a feedback signal which may be used to lock a
laser \cite{harvey}. In the circular basis, the phase difference,
$\phi$, between the left and right handed components can be
obtained from equations 2 and 3.  Evaluating the arctangent of the
ratio of the measured Stokes parameters $S_2/S_1$ yields a direct
measure of $\phi$. In practice, calculating the arctangent in real
time would require a digital signal processor.  However, for small
phase shifts, $\tan{\phi}$ $\approx$ $\sin{\phi}$ $\approx$ $\phi$
and $I_2$ provides a signal proportional to $\phi$. In
polarization spectroscopy \cite{pearman}, the small angle
approximation is valid, and a signal directly proportional to
$\phi$ can be obtained simply by measuring the single Stokes
component, $I_2$. The shape of the expected $I_2$ signal is given
by \cite{pearman},
\begin{eqnarray}
    I_2(x) \propto
    \Delta n(x) &=& \frac{\Delta \alpha_0}{k}\frac{x}{1 + x^2},
        \label{eq-birefringence}
\end{eqnarray}
where $x$ is the laser detuning from resonance in units of the
power broadened transition linewidth, $k$ is the (resonant)
wavenumber of the laser, $n(x)$ is the real part of the refractive
index of the vapor and $\Delta \alpha_0$ is the difference in
absorption at line center of the two circular components of the
probe.  This signal is dispersion shaped and free of a Doppler
broadened background (both left and right handed components
experience equal Doppler broadened absorptions and this is removed
in the subtraction).

The phase shift of equation \ref{eq-birefringence} depends only on
$\Delta \alpha_0$, the differential absorption at line center.
This difference originates from an anisotropy induced in atoms
which have interacted with the circularly polarized pump laser.
Atoms are optically pumped into one of the extreme magnetic
states, say $m_F$\,=\,+$F$. The differential absorption, $\Delta
\alpha$ of the two circular components of the probe field is
maximized in this case. In the 5$S_{1/2}$ $F$\,=\,2, $m_F$\,=\,2
ground state of $^{87}$Rb the difference in oscillator strengths
for $\sigma^+$ and $\sigma^-$ transitions is 15:1. The $\sigma^+$
component experiences enhanced absorption and the $\sigma^-$
decreased absorption with respect to the Doppler background. As
the Doppler broadened absorption is the same for both components
the differential absorption profile is given by the Lorentzian,
\begin{eqnarray}
    I_3(x) \propto
    \Delta \alpha(x) &=& \frac{\Delta
    \alpha_0}{1 + x^2},
    \label{eq-absorption}
\end{eqnarray}
which is related to equation \ref{eq-birefringence} by the
Kramers-Kronig dispersion relation \cite{demtroder}.  Equation
\ref{eq-absorption} is simply the difference in absorption of left
and right handed circular component (circular dichroism) which is
equivalent to $I_3$ of equation 4. Inspection of equations
\ref{eq-birefringence} and \ref{eq-absorption} reveals that their
quotient, $Q(x)$
\begin{eqnarray} Q (x) = \frac{I_2(x)}{I_3(x)} = \frac{x}{k},
\end{eqnarray}
provides a signal directly proportional to $x$ which does not
decay as 1/(1\,+\,$x^2$) at large detunings. In the polarization
locking of \cite{pearman}, $I_2$ is used as the error signal and a
controller provides feedback signal to the laser which drives this
to zero. $I_2(x)$ is linear for small $x$ but not at large
detunings where it approaches zero. Our locking method compensates
for this by dividing (\ref{eq-birefringence}) by
(\ref{eq-absorption}) which greatly extends the linear region of
the locking signal.

\begin{figure}[h]\centerline{\scalebox{1}
   {\includegraphics{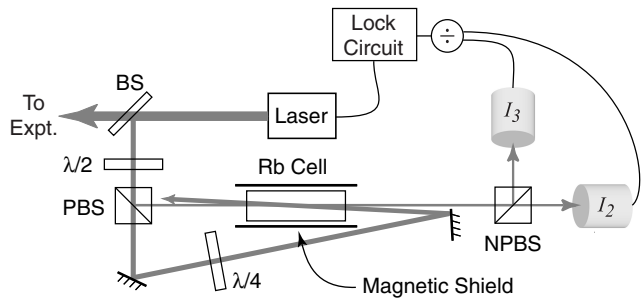}}}
   \caption{Experimental setup for polarization quotient locking.
   A small fraction of light is split off the laser output and
   sent into a polarization spectrometer.  The $I_2$ and $I_3$
   Stokes parameters of the probe light are measured and used to
   provide feedback to lock the laser. BS = beam splitter, PBS =
   polarizing BS, NPBS = 50/50 non-polarizing BS.}
   \label{fig:scheme}
\end{figure}

We have implemented this method spectroscopically and constructed
a simple control circuit for locking.  Figure \ref{fig:scheme} is
a schematic of our apparatus.  We choose for our experiments the
5$S_{1/2}$\,$F$\,=\,2 $\rightarrow$ $5P_{3/2}$\,$F'$\,=\,1,\,2,\,3
transition at 780.1nm in $^{87}$Rb. Light is provided by a 90\,mW
Rohm RLD78PZW1 diode laser in an external cavity configuration
similar to that described in \cite{arnold}.
Four photodetectors at the output of the optical apparatus are
connected to an analogue circuit which performs the necessary
subtractions and division in real time.  The output of this
device, either the polarization spectroscopy signal $I_2$, or the
quotient $I_2/I_3$, is fed into the laser lock circuit which
provides feedback to the laser frequency via the injection current
and piezoelectric transducer (PZT). Additionally, the lock circuit
has a differential input to which we can apply an offset voltage
which shifts the lock point away from zero volts.
\begin{figure}[h]\centerline{\scalebox{1}
   {\includegraphics{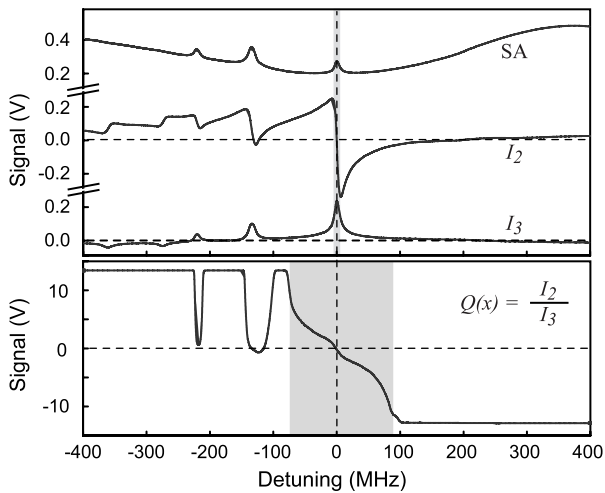}}}
   \caption{Spectra for the {$^{87}$Rb} 5S$_{1/2}$ $F$\,=\,2
   $\rightarrow$ 5P$_{3/2}$ $F'$\,=\,1,\,2,\,3 transition. Top trace
   is a regular saturated absorption spectrum (SA), the second is the $I_2$
   spectrum and third from the top is $I_3$. The trace below shows the
   quotient $Q(x)$.  The shaded regions indicate the monotonic capture
   range of $I_2$ (upper) and $Q(x)$ locking (lower).}
   \label{fig:spectra}
\end{figure}
Spectra were obtained as the laser frequency was scanned across
the transition by applying a triangular voltage ramp to the PZT.
Figure \ref{fig:spectra} shows the results of such a scan.  The
upper trace is a saturated absorption spectrum, followed by the
raw $I_2$ and $I_3$ scans. The lower trace shows the quotient,
$Q(x)$, obtained from the analogue divider. The horizontal scale
is obtained from the known energy level spacings of $^{87}$Rb in
zero magnetic field. These plots show that our method greatly
extends the capture range of the lock. The normalized gradients of
the quotient $I_2/I_3$\,=\,$S_2/S_3$ and $S_2$\,=\,$I_2/I_0$
spectrum are the same at $x$\,=\,0 but $Q(x)$ retains this value
at large detunings.

As $I_3$ approaches zero $Q(x)$ becomes sensitive to electronic
noise on the photodetector signals.  We overcome this by low-pass
filtering the output of the analogue divider with a 50\,kHz RC
filter.  Although, this limits the bandwidth of the lock loop, the
linewidth and long term stability of lasers locked with both $I_2$
and $I_2/I_3$ feedback signals were similar ($<$\,1\,MHz) as
determined by monitoring the error signals.

The linear region of the $I_2$ only locking signal extends over
20\,MHz. Our $Q(x)$ signal on the other hand, extends
monotonically over a 200\,MHz range as indicated by the shaded
regions in figure \ref{fig:spectra}. On the left, it is limited by
the proximity of the nearest crossover resonance, and on the right
only by the supply voltage to the analogue controller. This
increase in range helps in two important ways. Firstly, it means
the laser can sustain much larger perturbations without coming out
of lock and secondly, it makes it possible to apply large and
precise frequency steps without leaving the lock slope.

\begin{figure}[h]\centerline{\scalebox{1}
   {\includegraphics{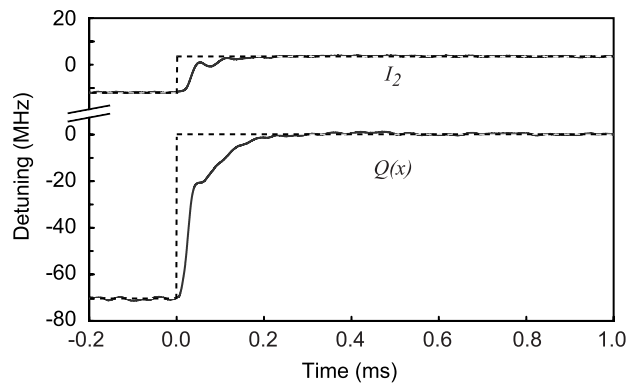}}}
   \caption{Step responses of a laser locked using $I_2$ and $Q(x)$
   locking after switching the DC lock point voltage at $t$\,=\,0.
   Solid lines show the measured error signal after switching and
   the dashed lines show the ideal response.  $Q(x)$ locking
   allows much larger jumps to be achieved without unlocking the
   laser.}
   \label{fig:step}
\end{figure}

In laser cooling and Bose-Einstein condensate experiments it is
often necessary to jump the frequency of a laser so that it can be
used for multiple tasks.  For example, a compressed
magneto-optical trap (MOT) or polarization cooling stage can be
achieved by shifting the MOT laser frequency from a detuning of
$\approx$\,-12\,MHz to $\approx$\,-50\,MHz.  Ideally this would
happen on a time scale of order 1\,ms. Acousto-optical modulators
can be used for this but a simpler method would be preferable.
With our scheme, such frequency steps can be achieved simply by
applying a DC offset voltage at the input to the lock circuit.
Figure \ref{fig:step} shows an example of this by plotting the
step responses of a laser locked using regular $I_2$ polarization
spectroscopy, and our quotient method. With $I_2$ locking, we can
repeatably jump the laser up to 15\,MHz in under 200\,$\mu$s.
Larger steps may take the laser off the lock slope and cause it to
relock at another zero crossing. Using quotient locking however,
we can jump from -70\,MHz back to resonance in approximately the
same time, making this technique very versatile for application to
laser cooling experiments. Similar jumps above resonance can also
be achieved.

In summary, we have seen that for a Lorentzian transition, the
ratio of the dispersion and absorption spectra yields a
measurement of detuning. Polarization spectroscopy obtains a
Doppler-free dispersion spectrum by measuring the unnormalized
Stokes parameter $I_2$.  A similar measurement can yield the
parameter $I_3$ which provides a Doppler-free measurement of the
differential absorption. Combining these produces a spectroscopic
measurement proportional to detuning ideal for use in laser
locking which extends the capture region of the lock by up to a
factor of ten. This results in a very robust lock and allows us to
make rapid and precise jumps of the laser frequency of up to
70\,MHz either side of resonance.

This work is supported by the Australian Research Council.


\begin{thebibliography}{}
%\begin{references}

\bibitem{demtroder}
W. Demtr\"{o}der, {\it Laser Spectroscopy} 2nd Edn., Springer,
Berlin (1998).

\bibitem{macadam}
K. B. MacAdam, A. Steinbach and C. Wieman, Am. J. Phys., {\bf 60},
1098 (1998).

\bibitem{white}
A. White, IEEE J Quantum Elctron., {\bf QE-1}, 349 (1965).

\bibitem{wieman}
C. Wieman and T. H\"{a}nsch, Phys. Rev. Lett., {\bf 36}, 1170
(1976).

\bibitem{hansch}
T. H\"{a}nsch B. Couillaud, Opt. Commun., {\bf 35}, 441 (1980).

\bibitem{pearman}
C. P. Pearman, C. S. Adams, S. G. Cox, P. F. Griffin, D. A. Smith
and I. G. Hughes, J. Phys. B: At. Mol. Opt. Phys, {\bf 35}, 5141
(2002).

\bibitem{harvey}
M. D. Harvey and A. G. White, Opt. Commun., {\bf 221}, 163 (2003).

\bibitem{bornwolf}
M. Born and E. WOlf, {\it Priciples of Optics} 7th Edn., Cambridge
University Press Berlin, Cambridge (1999).

\bibitem{arnold}
A. S. Arnold, S. J. Wilson and M. G. Boshier, Rev. Sci. Instrum.,
{\bf 69}, 1236 (1998).

\end{thebibliography}
\end{document}